\documentclass[letterpaper,12pt]{article}
\usepackage{graphics}
\begin{document}
\title{Quantum mechanics can be led by metrics in Minkowski's time-space world}
\author{Takashi Sakai}
\date{}

\pagestyle{plain}

\def\dfrac#1#2{{\displaystyle\frac{#1}{#2}}}

\maketitle
The Institute for Enzyme Research, The University of Tokushima\ 3-18-15 
Kuramoto-cho, Tokushima, 770-8503, Japan\\
\\
Present temporary address: Amgen Institute/ Ontario Cancer Institute, 
620 University Avenue, Suite 706, Toronto, Ontario, M5G 2C1, Canada\\
\\
Correspondence should be addressed to the author (e-mail:
 sakai@ier.tokushima-u.ac.jp, Tel.:+81-88-633-7430, Fax:+81-88-633-7431)\\
\begin{abstract}
A body of theory is completely different between relativity theory and 
quantum mechanics.  Most targeting physical phenomena are different between 
them as well.  Despite that both theories describe our time-space world, no 
one has succeeded in arriving at one of these theories by using the other 
one, deductively.  Fusion of these two theories produced quantum field theory.  However this theory impresses us as a theory that was made by merging the two 
theories by force.  So this theory cannot indicate the reason why both 
relativity theory and quantum mechanics exist in this world at the same 
time.  Here I show you that quantum mechanics can be led by the equation of 
Minkowski's metrics, which stands up in the relativistic world of space and 
time.  This is the first report proving mathematically that quantum 
mechanics can be born reductively in our relativistic world.
\end{abstract}
\vspace{\baselineskip}
The strong success of quantum mechanics is not surprising considering its 
much greater degree of efficacy, on an atomic scale, compared to any other 
theory.  However, it has also been difficult to understand the fundamental 
implication of quantum mechanics because of its specificity from the 
structure of the theory including wave-particle duality\cite{Fynman}.  
We are able to see such specificity from the aspect that the theory has 
borne lots of 
paradoxical questions.  On the other hand, the theory includes another 
important aspect: that it seems to be incompatible with aspects of 
relativity theory that exert great validity against our macroscopic scale.  
Here I would like to propose some novel techniques and ideas in order to 
understand the relationship between these two theories and to solve the 
paradoxical questions of quantum mechanics, such as wave-particle duality 
or the measurement problem.  Here it is shown that quantum mechanics can 
be led from the equation in Minkowski's space-time world.\\

The differential distance $ds$ in Minkowski's space-time world (Minkowski's 
metrics) is given by\\
\begin{equation}
(ds)^2=-(c\,dt)^2+(dx)^2+(dy)^2+(dz)^2
\end{equation}
For our simplified spacetime, with just one space dimension, this reduces to\\
\begin{equation}
(ds)^2=-(c\,dt)^2+(dx)^2
\end{equation}
This can also be rewritten by
\begin{eqnarray}
(ds)^2 &=& -(c\,dt)^2(\cos^2\theta_t-(i\sin\theta_t)^2)
+(dx)^2(\cos^2\theta_x-(i\sin\theta_x)^2)\nonumber \\
&=& -(c\,dt\cos\theta_t,ic\,dt\sin\theta_t)
(c\,dt\cos\theta_t,-ic\,dt\sin\theta_t) \nonumber \\
& & {} +(dx\cos\theta_x,i\,dx\sin\theta_x)
(dx\cos\theta_x,-i\,dx\sin\theta_x) \nonumber \\
&=& -d\psi_t\cdot d\psi_t^*+d\psi_x\cdot d\psi_x^*
\end{eqnarray}
\begin{eqnarray*}
i = \sqrt{-1}\\
d\psi (\theta) &=& dx (\cos\theta,i\sin\theta)\\
d\psi^*(\theta) &=& dx (\cos\theta,-i\sin\theta) = d\psi(-\theta)
\end{eqnarray*}
These $d\psi$ and $d\psi^*$ are expressed as the coordinates on a 
circumference in a complex space of real and imaginary axis (Figure 1).  
I will discuss later concerning this complex space.  First of all, you 
consider a complex world consisting with an imaginary space lying at right 
angles to a real space for either its 
time dimension or its space dimension.  Then you think about a circle 
consisting of a minute segment ($dx$) as a radius (Figure 1).
The ``$d\psi (\theta) = dx (\cos \theta,i\sin \theta)$ " or the 
``$d\psi^* (\theta) = dx (\cos \theta,-i\sin \theta)$ " shown above 
represents points on the circumference.  If this condition $\psi$ is 
involved in a quantum behavior, $\theta$ must be a function of the 
coordinates of the real space, $x$.  When you consider the existence 
of a particle in this minute region $\Delta x$, the condition must 
have a feature of a wave in the imaginary space.  If you use $\lambda$ 
for its wavelength, you can substitute $\dfrac{2\pi}{\lambda}x$ for 
$\theta$.  So if you write the particle's condition as the function of 
$x$ like $\psi(x)$, the condition in the minute region $d\psi(x)$ is 
written as $d\psi(x)=dx(i\sin\dfrac{2\pi}{\lambda}x,\cos\dfrac{2\pi}
{\lambda}x)$.  This can be rewritten as $\dfrac{d\psi}{dx}=
(i\sin\dfrac{2\pi}{\lambda}x,\cos\dfrac{2\pi}{\lambda}x)$.  And this 
leads to $\psi=-i\dfrac{\lambda}{2\pi}(i\sin\dfrac{2\pi}
{\lambda}x,\cos\dfrac{2\pi}{\lambda}x)=-i\dfrac{\lambda}{2\pi}
\dfrac{d\psi}{dx}$.  Assuming that the $\lambda$ is equivalent to the 
de Broglie wavelength,\cite{Broglie} the relationship between the $\lambda$ 
and the momentum of the particle ($P$) can be described as $\dfrac{\lambda}
{2\pi}=\dfrac{\hbar}{P}$.  If you substitute this, you get the equation 
shown below.\\
\begin{equation}
P\psi = -i\hbar\frac{\partial \psi}{\partial x}
\end{equation}
Next, you think about time dimension as well.  You combine the equation 
$d\psi(t)=dt(i\sin\theta,\cos\theta)$ and $\theta=-2\pi\nu t$.\\
Then you get $\dfrac{d\psi}{dt}=c(i\sin(-2\pi\nu t),\cos(-2\pi\nu t))$.  
One of the simplest solutions of the equation is $\psi(t)=i\dfrac{c}
{2\pi\nu}(i\sin(-2\pi\nu t),\cos(-2\pi\nu t))=i\dfrac{1}{2\pi\nu}
\dfrac{d\psi}{dt}$.  By substituting $\nu=\dfrac{E}{h}$ into this 
equation, you get
\begin{equation}
E\psi(t)=i\hbar\frac{d\psi}{dt}
\end{equation}
Equivalent results can be obtained concerning $\psi^*(x)$ and $\psi^*(t)$ 
as well.  So, all the results can be put together with the simple 
equations shown below.\begin{equation}
(ds)^2=g_{ij}\,dx_i\,dx_j=g_{ij}\,d\psi_i\,d\psi^*_j
\end{equation}
\begin{center}
($g_{ij}$: metric tensor)\\
\end{center}
\begin{equation}
\psi^*(x)=\psi(-x)=-\frac{\hbar}{P}(i\cos\frac{P}{\hbar}x,\sin\frac{P}
{\hbar}x)
\end{equation}
\\
Here I succeeded in demonstrating that quantum mechanics can be led by 
Minkowski's space-time world by assuming the existence of an imaginary 
space lying at right angles to a real space for either its time dimension 
or its space dimension.  You can also see from the equation of $\psi$ or 
$\psi^*$ that the quantum mechanical area size of 
a particle is restricted in $\dfrac{\hbar}{P}$, in which the particle 
exists as two waves written by complex numbers ($\psi$ and $\psi^*$).  
This means that any matter produces a quantum field by its momentum, 
and that the size of the field is restricted by the momentum.  In 
addition, both of the quantum mechanical functions led by two wave 
functions $\psi$ and $\psi^*$ are equivalent, so these quantum 
mechanical features are undeniably with each other.  The conclusion is 
that any particle moves as two quantum mechanical waves of $\psi$ and 
$\psi^*$ in its specific small areas.  On the other hand, if you write 
the condition of a particle without the idea of an imaginary space, you 
cannot see such quantum mechanical features.  Here you get a solution 
against the measurement problem of quantum theories.  The measurement 
itself is equivalent to denying the existence of an imaginary space 
because we cannot measure any movement in imaginary spaces.  So if you 
describe the movement according to the measurement, you never reach the 
functions of $\psi$ and $\psi^*$.  So you can only measure the space by 
$(dx)^2=g_{ij}\,dx_i\,dx_j$, which is an incomplete equation of the 
equation (6).\\
Next, I discuss the meaning of the complex world of real and imaginary axes.  
The imaginary axis lying at right angles to a real axis (Figure 1) can be 
interpreted as an axis of the fifth dimension, which is different from our 
four-dimensional world.  However, more simple and easy alternative idea may 
be to interpret it as a time axis.  On the contrary, an imaginary axis against 
time axis can be interpreted as an axis of the fifth or sixth dimensions.  
However, the more simple and easy way is to interpret it as a space axis.  
For instance, you put a space axis and an imaginary axis lying at right 
angles to the space axis as $x$ and $iy$, respectively ($i=\sqrt{-1}$), 
metric tensor $(ds)^2$ is given by\\
\begin{equation}
(ds)^2=(dx)^2+(idy)^2=(dx)^2-(dy)^2
\end{equation}
This indicates that the $iy$ axis is in a time dimension.  In addition, the 
equation (3) indicates that both the space and time axes can be treated as 
the complex of real and imaginary axes in a minute region.  And this 
operation can lead quantum mechanics as I showed.  These facts lead us to 
imagine that the essence of quantum mechanics lies in the transformation 
between space and time regions in the world of space and time.
\begin{tabular}{c}
\hline
\end{tabular}

\begin{figure}[htbp]
	\includegraphics*{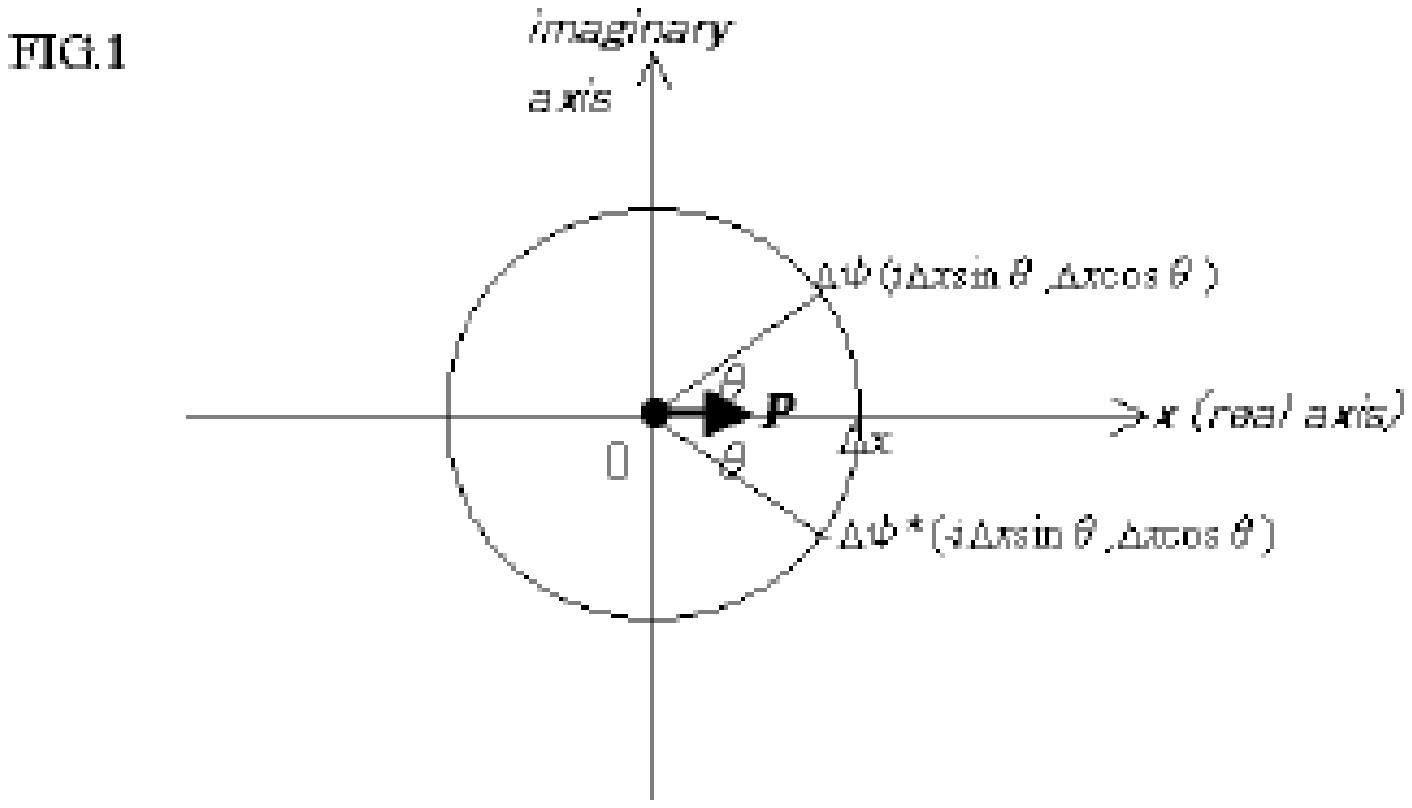}
	\caption{Particles exist in complex world}{I hypothesize that an 
	imaginary world exists against our real world of time and space.  
	The equation (3) indicates that the axes of the imaginary world lie 
	at right angles to the axes of the real world.  You consider a 
	particle, which is moving with momentum $P$ at a moment.  This 
	particle is supposed to have an uncertain region ($\Delta x$).  
	And this uncertain region can be composed of two different 
	conditions ($\Delta\psi$ and $\Delta\psi^*$) according to the 
	equation (3).}
	\label{fig:Figure1.eps}
\end{figure}
\end{document}